\documentclass[twocolumn]{aastex61}

\usepackage{amsmath}
\usepackage[utf8]{inputenc}
\usepackage[]{graphicx}
\usepackage[varg]{txfonts}
\usepackage{enumerate}
\usepackage{float}
\usepackage{xspace}
\usepackage{balance}
\usepackage{upgreek}
\usepackage{array}

\received{--}
\revised{--}
\accepted{--}
\submitjournal{ApJ}
\shorttitle{Alternate Approach to Measure sSFR at $2<z<7$}
\shortauthors{I.~Davidzon et al.}

\begin{document}

\title{An alternate approach to measure specific star formation rates at $2<z<7$}

\correspondingauthor{I.~Davidzon}
\email{iary@ipac.caltech.edu}

\author{Iary Davidzon}
\affil{IPAC, Mail Code 314-6, California Institute of Technology, 1200 East California Boulevard, Pasadena, CA 91125, USA}

\author{Olivier Ilbert}
\affil{Aix Marseille Universit\'e, CNRS, LAM (Laboratoire d'Astrophysique de Marseille) UMR 7326, 13388, Marseille, France}

\author{Andreas L.~Faisst}
\affil{IPAC, Mail Code 314-6, California Institute of Technology, 1200 East California Boulevard, Pasadena, CA 91125, USA}

\author{Martin Sparre}
\affil{Institut f\"ur Physik und Astronomie, Universit\"at Potsdam, Karl-Liebknecht-Str.\,24/25, 14476 Golm, Germany}

\author{Peter L.~Capak}
\affil{IPAC, Mail Code 314-6, California Institute of Technology, 1200 East California Boulevard, Pasadena, CA 91125, USA}

\begin{abstract}
We trace the specific star formation rate (sSFR) of massive star-forming galaxies ($\gtrsim\!10^{10}\,\mathcal{M}_\odot$) from $z\sim2$ to 7. Our method is substantially different from previous analyses, as it does not rely on direct estimates of star formation rate, but on the differential evolution of the galaxy stellar mass function (SMF). We show the reliability of this approach by means of semi-analytical and hydrodynamical cosmological simulations. We then apply it to real data, using the  SMFs derived in the COSMOS and CANDELS fields. We find that the $\mathrm{sSFR}$ is proportional to $(1+z)^{1.1\pm0.2}$ at $z>2$, in agreement with other observations but in tension with the steeper evolution predicted by simulations from $z\sim4$ to 2. We investigate the impact of several sources of observational bias, which however cannot account for this discrepancy. Although the SMF of high-redshift galaxies is still affected by significant errors, we show that future large-area surveys will substantially reduce them, making our method  an effective tool to probe the massive end of the main sequence of star-forming galaxies. 
\end{abstract}


%%%%%%%%%%%% BEGIN PAPER %%%%%%%%%

\section{Introduction}

In less than 4 Gyr, between the Epoch of Reionization ($z\sim8$) and the Cosmic Noon ($z\sim2$), galaxies build almost half of the local universe's stellar content  \citep{Madau2014}.  
Key quantities to describe such a growth are galaxy stellar mass ($\mathcal{M}$) and star formation rate (SFR), whose ratio is the galaxy specific star formation rate 
($\mathrm{sSFR}\equiv\mathrm{SFR}/\mathcal{M}$, i.e.~the rate of mass doubling of a galaxy). 

The sSFR of star-forming galaxies reflects their ``main sequence'' (MS) distribution \citep{Noeske2007}. Its evolution is a primary constraint both on processes that govern stellar mass accretion and the ones responsible for its cessation (the so-called ``quenching'' mechanisms).  
For instance, \citet{Renzini2016} used analytical fits to sSFR($\mathcal{M},z$) and $\Psi(z)$ -- the cosmic SFR density -- to predict the galaxy quenching rate as a function of redshift \citep[see also][for  a similar approach]{Peng2010,Boissier2010}. 
However this kind of analyses are still affected by significant uncertainties: at present there is no full concordance among the various sSFR measurements, especially at high redshift. 

\begin{figure}
\includegraphics[width=0.99\columnwidth]{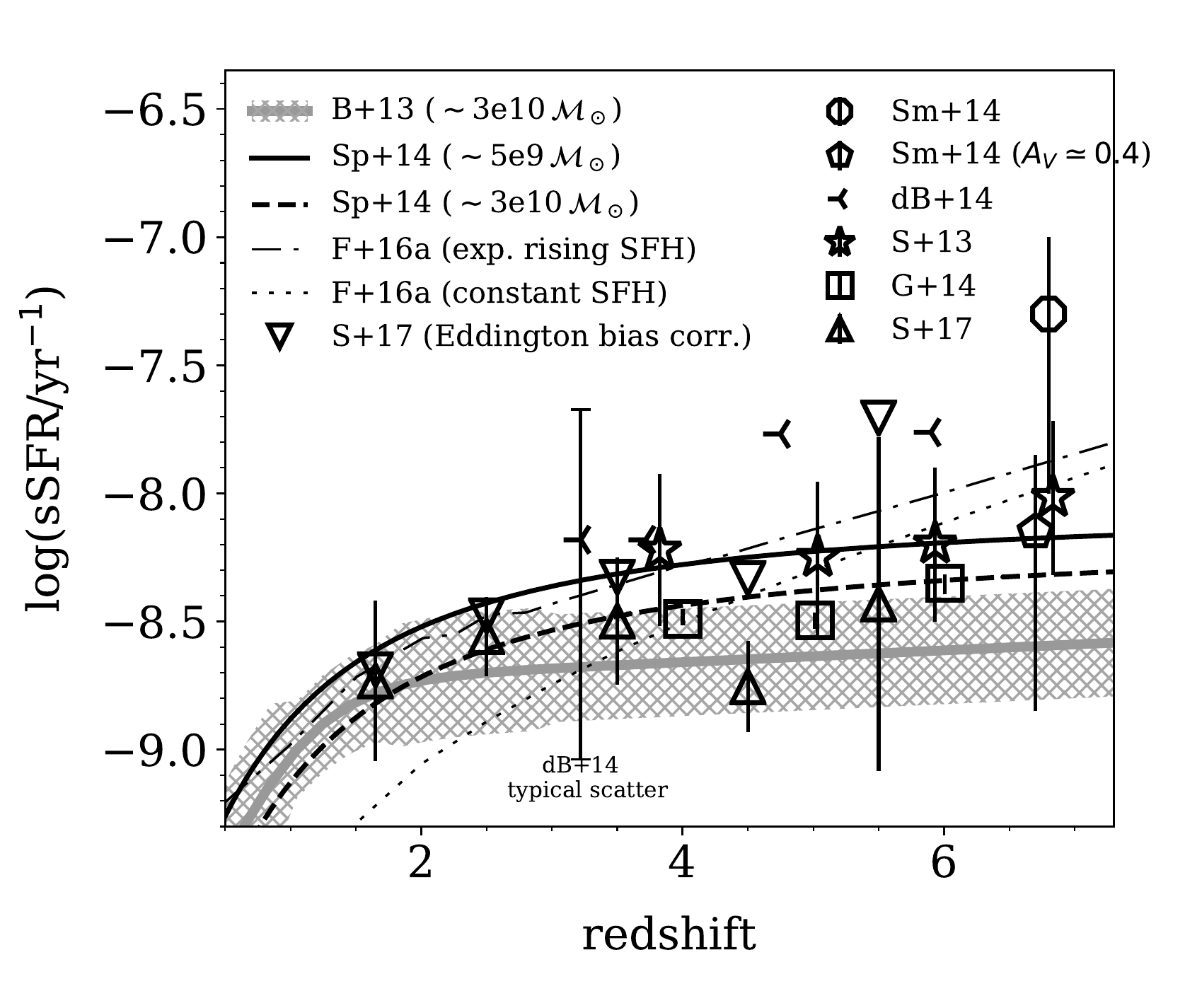}
\caption{Estimates of sSFR($\mathcal{M},z$) from the literature, derived at $\mathcal{M}=5\!-\!30\times10^{9}\,\mathcal{M}_\odot$ (depending on the study). The gray line with hatched area is the fit to data from studies published between 2007 and 2013 \citep[][]{Behroozi2013b}. Another fit to the updated compilation of \citet[][]{Speagle2014} is shown by a solid (dashed) line for sSFR at $\log(\mathcal{M/M}_\odot)\simeq9.7$ ($10.5$). 
Other symbols (see legend) show individual measurements from  \citet[][]{Stark2013},  \citet[][]{Gonzalez2014}, 
\citet[][]{deBarros2014b}, 
\citet[][with different priors on dust extinction]{Smit2014}, \citet[][different star formation histories]{Faisst2016a} and \citet[][also including the Eddington bias correction]{Santini2017}.   
Not to compromise the readability of the figure, we show the error bar of only one point of \citet{deBarros2014b}; this is the 68\% confidence limit derived from the whole probability distribution function  of their SED models. 
} 
\label{fig:ssfr_lit}
\end{figure}

Figure~\ref{fig:ssfr_lit} summarizes the state of the art in this context and highlights the substantial discrepancies in the measurements of the different studies. 
In this Figure, as well as in the rest of the paper, we refer to the sSFR$(z)$ computed at a constant stellar mass.   

Pioneering studies at $z>3$ found a plateau in the sSFR evolution \citep[e.g.][]{Stark2009,Gonzalez2010}, in tension with theoretical predictions \citep[see][]{Weinmann2011}. 
In those papers $\mathcal{M}$ and SFR were derived by fitting their photometry with synthetic spectral energy distributions (SEDs) neglecting nebular emission contamination in the broad-band filters. This introduced a substantial bias in the sSFR estimates, as shown in Fig.~\ref{fig:ssfr_lit} where a compilation of those studies \citep[][]{Behroozi2013b} is compared to more recent work.  
The latter, after accounting for optical emission lines in the SED fitting, find an increasing  sSFR$(z)$ from $z=3$ to at least 7 \citep[][]{Stark2013,Gonzalez2014,deBarros2014b}.  
Another example of SED fitting systematics is shown in 
\citet{deBarros2014b}: their results change by an order of magnitude with varying assumption on star formation history (SFH), age, and metallicity. Moreover, when   
 $\mathcal{M}$ and SFR are both derived through SED fitting, resulting biases are inter-connected and more difficult to correct.

Other studies use a different technique based on emission line contamination. They start from the color excess in broad-band filters to estimate H$\alpha$ equivalent width (EW), which is a good  proxy for the sSFR.  
This novel approach does not rely on classical SED fitting recipes, even though it also makes use stellar population synthesis (SPS) models and needs assumptions on the SFH. Depending on the photometric baseline, it can yield results at $z=4\!-\!5$     \citep{Shim2011,Rasappu2016} or over a larger redshift range \citep[$1<z<6$,][]{Faisst2016a,MarmolQueralto2016}.

Figure \ref{fig:ssfr_lit} marks other two caveats relevant for any sSFR estimator, namely  the dust correction \citep[see discussion in][]{Pannella2015} and the observational biases \citep[e.g.~the Eddington bias,][]{Eddington1913}. Both have a significant impact especially at high redshift \citep[][]{Smit2014,Santini2017}.  
Furthermore, diverging assumptions of the contamination of [N\textsc{II}] to H$\alpha$ affecting the SFR measurements from low-resolution spectroscopic surveys can lead to differences in the sSFR determinations (Faisst et al.~in preparation).
In addition, we note that -- since the sSFR at fixed $z$ depends on stellar mass \citep[e.g.,][]{Whitaker2014} -- the scatter  among the various estimates shown in Fig.~\ref{fig:ssfr_lit} is also due to the fact that the sSFR is derived at different masses (between $5\times10^{9}$ and $3\times10^{10}\,\mathcal{M}_\odot$).

As a consequence, despite the improvements in observations, the  sSFR($z$) function is still matter of debate. Several authors find an increase by a factor $\sim\!5$  across $3<z<7$ \citep[e.g.][]{Stark2013,deBarros2014b,Salmon2015,Faisst2016a} while others observe a flatter evolution \citep[e.g.][]{Gonzalez2014,Heinis2014,Tasca2015,MarmolQueralto2016}.
Different slopes imply discordant scenarios of galaxy evolution, with respect to gas accretion, stellar mass assembly, and quenching time-scales \citep{Weinmann2011}. 
Some discrepancy with simulations remains, especially at $1<z<3$ \citep[e.g.][]{Dave2016}.  

In this paper we provide a new sSFR constraint for star-forming galaxies up to $z\sim7$, through a novel approach based on the evolution of their stellar mass function (SMF).  
Such a method, described in Sect.~\ref{sect:method}, offers a complementary point of view with respect to previous work. In fact, we rely on integrated quantities only (stellar mass and cumulative galaxy number density) without any direct SFR assessment. 
We demonstrate the validity of our approach by the use of cosmological simulations (Sect.~\ref{sect:simu}) before applying the method to real data (Sect.~\ref{sect:results}). 
The results are discussed in Sect.~\ref{sect:discussion} and then we conclude in Sect.~\ref{sect:conclusions}. 

We assume a $\Lambda$CDM  cosmology with  $\Omega_\mathrm{m}=0.3$, $\Omega_\Lambda=0.7$, and $H_0=70\,\mathrm{km}\,\mathrm{s}^{-1}\,\mathrm{Mpc}^{-1}$. The initial mass function (IMF) used as a reference is \citet{Chabrier2003}. We assume that the SMF shape is well described by a \citet{Schechter1976} function, or a combination of two Schechter functions (with the same $\mathcal{M}_\star$ parameter) at $z<3$.

%%%%%%%%%%%%%%%% FIG 1 %%%%%%%%%%%%%%%%%%%
%%%%%%%%%%%%%%%%%%%%%%%%%%%%%%%%%%%%%%%%%%%%%%%%%

%%%%%%%%%%%%%%%%% SECTION METHOD

\section{Method}
\label{sect:method}

\subsection{Motivations}
\label{sect:method1}

Before describing our method we highlight some of the main difficulties encountered in previous sSFR studies:
\begin{enumerate}[i)]
\itemsep0em
\item collecting a fully representative sample of star-forming galaxies, in a given mass range;   
\item implementing realistic star formation histories in the SPS models;
\item defining physically motivated SED-fitting parameters and priors.
\end{enumerate}
Limitations due to sample incompleteness are discussed e.g., in \citet{deBarros2014b} and 
\citet{Speagle2014}. 
Furthermore, recent work has reevaluated the fraction of star-forming galaxies at high-$z$ that are strongly enshrouded by dust \citep[e.g.][]{Casey2014a,Mancini2015}. This population may be  missing in Lyman break galaxy (LBG)  selections \citep[see discussion in][]{Capak2015}. 
Argument ii) has a limited impact on stellar mass estimates 
\citep{Santini2015} whereas the SFR   
is extremely sensitive to the SFH details (such as secondary bursts and star formation ``frostings'').  
Concerning the third issue, examples of critical parameters are stellar metallicity,  dust reddening, or the EW  of nebular emission lines. Modifying their parametrization, or the range of allowed values, can produce significant differences both in  $\mathcal{M}$ and SFR estimates   \citep[][]{Conroy2009b,Mitchell2013,Stefanon2015} as well as in their covariance matrix. The systematic effects inherited by the sSFR are discussed e.g.~in \citet{Stark2013} and \citet{deBarros2014b}. 

To circumvent these limitations as well as possible, we follow the semi-empirical approach described in 
\citet[][]{Ilbert2013}.  The keystone of our method  is the SMF of star-forming and quiescent galaxies. Their evolution from high ($z_i$) to low ($z_f<z_i$) redshift brings information on the stellar mass assembly \citep[see also][]{Wilkins2008}. We perform our analysis in the COSMOS and CANDELS fields (\citealp{Scoville2007} and \citealp{Grogin2011}) because their latest observations allows  to probe the SMF with unprecedented accuracy and to higher redshifts.  

\subsection{Evolution of a single galaxy}
\label{sect:method2}

The SMF evolution from  $z_i$ (cosmic time $t_i$) to $z_f$ (i.e., $t_f>t_i$) can be modeled starting from the growth of individual galaxies. 
Broadly speaking, galaxy stellar mass accretion occurs through two channels: \textit{in situ} star formation and galaxy merging. 
If we consider only the former process, the stellar mass of a galaxy at $z_f$  can be written as
\begin{equation}
\mathcal{M}(t_f) = \mathcal{M}(t_i) + \int_{t_i}^{t_f}  \mathit{SFR}(t) \left[1-f_\mathrm{return} \right]\,\mathrm{d}t,  
\label{eq:delta_m}
\end{equation} 
where the integral of the SFR  accounts for the mass re-ejected into the interstellar medium via stellar winds and supernovae ($f_\mathrm{return}$). This can be taken as the instantaneous return fraction ($f_\mathrm{return}=0.43$ for a Chabrier IMF) or defined as a function of time as in \citet{Behroozi2013b}: 
\begin{equation}
f_\mathrm{return} = 0.05\ln\left[1+\frac{t_f-t}{1.4\times10^6\,\mathrm{yr}}\right]. 
\label{eq:f_return}
\end{equation}
One can simplify Eq.~(\ref{eq:delta_m}) 
by assuming that SFR($t$) is constant across $\Delta t\equiv t_f-t_i$. 
This is a fair approximation of the average SFH in the high-$z$ universe, at least on short time-scales like the $\Delta t$ steps that we consider (see below). 
A better fit may be a rising function, e.g.~$\mathrm{SFR}\propto t^{\gamma}$ with $\gamma\sim1.4\!-\!4$ \citep{Papovich2011,Behroozi2013b}. 
We discuss 
the implications of this choice in Section \ref{sect:simu2}.

In addition, the fractional mass increase via mergers is
\begin{equation}
f_\mathrm{merg} = \int_{t_i}^{t_f} \frac{\dot{\mathcal{M}}_\mathrm{merg}}{\mathcal{M}(t_i)} \,\mathrm{d}t = \int_{t_i}^{t_f}  \frac{\mathcal{R}(t)}{\bar{\mu}} \,\mathrm{d}t,  
\label{eq:m_merg}
\end{equation}
where $\dot{\mathcal{M}}_\mathrm{merg}$ is the merger-driven stellar mass accretion rate, which can be written as a function of $\mathcal{R}$ (the merger rate) and $\bar{\mu}$ (the average stellar mass ratio between the target galaxy and the accreted satellite). 
%In general $\mathcal{R}$ depends also on $\mathcal{M}$, but we drop this dependency because we operate at a fixed mass. 

Equation~(\ref{eq:m_merg}) takes into account only the stellar mass already formed \textit{ex situ} and accreted onto the given galaxy, neglecting possible bursts of star formation triggered by the galaxy-galaxy interaction. 
The latter is a well-established phenomenon in the local universe whereas it is less clear whether at higher redshift mergers can induce significant starburst episodes.  
In the standard hierarchical clustering scenario, the SFR enhancement per single merger is expected to increase with redshift as it is more likely that galaxy pairs have comparable mass \citep[a condition for efficiently triggering starbursts,][]{Cox2008};  
the fraction  of  destabilized  gas  is also larger at earlier epochs. However, recent hydrodynamical pc-resolution simulations \citep{Fensch2017} show that such a large gas fraction (and gas clumpiness) does result in strong inflows and turbulence already in isolated objects, thus the interaction-induced star formation causes only a mild SFR increase (a factor $2\!-\!3$) over short timescales ($\sim\!50$\,Myr). These findings are in agreement with  other high-resolution prototypes of high-$z$ gas-rich mergers \citep{Hopkins2013,Perret2014} but also with previous analyses \citep{Cox2008}. In another hydrodynamical simulation, this time with cosmological size, \citep{Martin2017} find that the average enhancement due to either major or minor mergers is about 35\% at $z\sim3$.   
We also modify the model assuming $\times2$ star formation increase  \citep{Robaina2009} over  100\,Myr, finding negligible changes in our results.
Therefore we decided not to include merger-driven starbursts in Equation (\ref{eq:m_merg}).

We fix $\dot{\mathcal{M}}_\mathrm{merg}$ to be equal to $5\times10^{9}\,\mathcal{M}_\odot\,\mathrm{Gyr}^{-1}$, according to \citet{Man2016}. The authors derive this value for galaxies with $\log(\mathcal{M/M}_\odot)>10.8$ at $0.5<z<2.5$  including both major and minor mergers, for which 
they observe  $\mathcal{R}\sim0.1\!-\!0.2$ and  $\mathcal{R}\lesssim0.1$  respectively. These values are also in good agreement with simulations  
\citep{Hopkins2010}. 
We extrapolate \citeauthor{Man2016} results also at $z>2.5$ as they are consistent with the latest studies at higher redshift. 
For example observations in COSMOS and CANDELS indicate that for galaxies with $\mathcal{M}>10^{10}\,\mathcal{M}_\odot$ the major merger rate is $\mathcal{R}\leqslant0.1\,\mathrm{Gyr}^{-1}$  up to $z\sim3.5$, with an extrapolated trend towards higher $z$ that is nearly flat  \citep[][Duncan et al.~in preparation]{Mundy2017}. Observations with the Multi-Unit Spectroscopic Explorer (MUSE) in the \emph{Hubble} Ultra Deep Field also find the same trend, with the major merger fraction that peaks at 20\% at $z=2\!-\!3$ and then decreases towards $z=6$ 
\citep{Ventou2017}. Equation (\ref{eq:m_merg}) will be re-tuned  when additional data come out. We show the impact of the assumptions about $f_\mathrm{return}$ and $f_\mathrm{merg}$ in  Section~\ref{sect:simu2}.

Eventually, with the approximation of constant SFR,  the relation between the logarithmic increase $\Delta \log\!\mathcal{M}\equiv \log[\mathcal{M}(t_f)/\mathcal{M}(t_i)]$ and the galaxy sSFR is
\begin{equation}
\mathit{sSFR}(\mathcal{M}_i,t_i)  
= \frac{10^{\Delta \log\!\mathcal{M}} -1 - f_\mathrm{merg}}{\Delta t -\int_{t_i}^{t_f} f_\mathrm{return} \,\mathrm{d}t}, 
\label{eq:ssfr_ev} 
\end{equation}
where we define $\mathcal{M}_i\equiv\mathcal{M}(t_i)$ for sake of clarity (i.e., the sSFR estimates we will show hereafter correspond to the initial redshift bin $z_i$).    
We note that $\Delta \log\!\mathcal{M}$ is the total stellar mass increase observed in a galaxy. 
To recover the net amount of stars formed \textit{in situ}, and then the sSFR, this quantity must be corrected for mergers  (Equation~\ref{eq:m_merg}) and stellar mass loss  (Equation~\ref{eq:f_return}) over the time interval   $\Delta t$.

\subsection{Matching sSFR($z$)  to the SMF evolution}
\label{sect:method3}

\begin{figure}
\centering
\includegraphics[width=0.99\columnwidth]{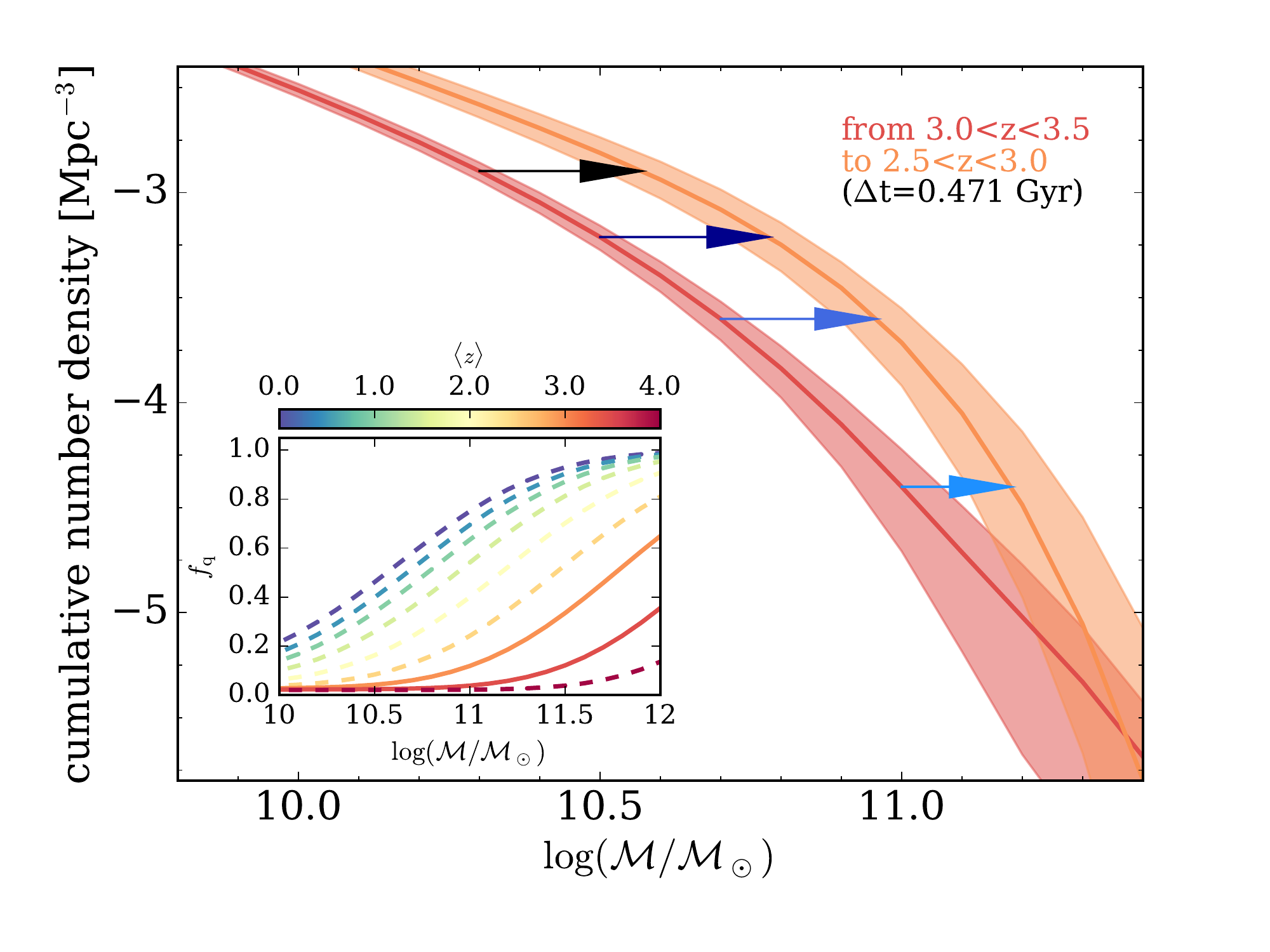}
\caption{An example of abundance matching at constant cumulative galaxy number density, by integrating 
the star-forming SMFs measured in COSMOS2015 \citep[from][]{Davidzon2017} at  $3<z<3.5$ and $2.5<z<3$  (red and orange line respectively, with shaded areas encompassing $1\sigma$ uncertainty). The four  arrows trace the growth of galaxies that in the higher $z$-bin have $\mathcal{M}\sim2$, 3, 5, and $10\times10^{10}\,\mathcal{M}_\odot$.  The number density of star-forming galaxies in the lower $z$-bin is corrected for recent quenching by removing the fraction of quiescent galaxy  ($f_\mathrm{q}$). The $f_\mathrm{q}(\mathcal{M},z)$ evolution is shown in the inset, as it results from COSMOS2015 data \citep[][the two $z$-bins of interest are highlighted with solid lines]{Bethermin2017}.  
}
\label{fig:cumu_smf}
\end{figure}

In order to apply Equation~(\ref{eq:ssfr_ev}) one needs an estimate of $\Delta \log\!\mathcal{M}$. The formalism introduced in  Section~\ref{sect:method2} describes the average  growth in stellar mass, therefore we can look to the galaxy ensemble as encoded in the SMF.   
At a given stellar mass, we link  star-forming galaxies at $z\sim z_i$   to their descendants at $z_f$ by tracking their cumulative number density \citep[$\rho_N$, see][]{vanDokkum2010,Behroozi2013c,Torrey2015}. 
This is obtained from the integral of the star-forming SMF. However, some galaxies in the initial $z$-bin may quench their star formation before $z_f$. For this reason, the star-forming $\rho_N(>\mathcal{M}_i)$ has to be corrected for  the increased number density of quiescent galaxies \citep[see][]{Ilbert2013}. Their fraction, as a function of $z$ and $\mathcal{M}$, can be derived from the quiescent SMF  \citep[inset in Figure \ref{fig:cumu_smf}, and also Figure 16 of][]{Faisst2017a}.

We connect galaxies at constant $\rho_N(>\mathcal{M}_i)$ from $z_i$ to $z_f$ \citep[][]{vanDokkum2010}.   
 Figure \ref{fig:cumu_smf} shows an example of such a procedure, using the SMFs observed in the COSMOS field at $z_i\simeq3.25$ and $z_f\simeq2.75$. 
 Arrows in the figure show $\Delta\log\!\mathcal{M}$ for a constant $\rho_N(>\mathcal{M}_i)$ evolution at different stellar masses.  We choose   $10^{10}<\mathcal{M}_i<10^{11}$, as this is the range where the SMF is well constrained by our data across the whole redshift range.

Then we repeat the procedure accounting for density evolution in the abundance matching. 
When connecting galaxies in the cumulative SMF to their descendants in the next $z$-bin, their  
$\mathcal{M}$ rank order may be different from the progenitors because of  mergers and SFR scatter   \citep[e.g.,][]{Leja2013}. In this case the  merging events that must be taken into account are not only those involving a target  galaxy (Section \ref{sect:method2}): the cumulative distribution is also modified  by mergers between galaxies in lower mass bins that are promoted in the one for which we  derive the sSFR. 
The SFR scatter also modifies the galaxy ranking  in the abundance matching, as some low-mass galaxies can grow faster than others with higher mass.\footnote{
We also note that the intrinsic scatter in the MS \citep[$\sim\!0.2$\,dex,][]{Speagle2014} and the small fraction of  $\mathcal{M}>10^{10}\,\mathcal{M}_\odot$ outliers \citep[$\lesssim10\%$,][]{Rodighiero2011,Caputi2017}
indicate that the SFR scatter does not bias the median sSFR we want to derive.}
To correct the abundance matching for these effects  we use the model provided by \citet{Torrey2015}.\footnote{
 \url{https://github.com/ptorrey/torrey_cmf}}
The authors track the cumulative SMF at different epochs using the merger trees of the Illustris simulation \citep[][]{Genel2014,Vogelsberger2014a,Vogelsberger2014b} and they fit the resulting number density evolution as a function of $z$ and $\mathcal{M}$. 
Following their recipe the  $\rho_N$ threshold of galaxies at $t_f$ is slightly higher than their progenitors. However, as we will show in Section \ref{sect:simu1}, this 
modification is a second-order effect that does not change any of our results \citep[similarly to ][]{Salmon2015,Stefanon2017b}.

Concerning the issues listed in Section \ref{sect:method1}, we emphasize the following advantages of our method:
%%%%
\begin{enumerate}[i)]
\item The SMF is corrected for incompleteness \citep[in our case through the $1/V_\mathrm{max}$ method,][]{Schmidt1968}.
\item Even though the stellar mass of some peculiar galaxy class may strongly depend on the SFH \citep{Michalowski2014}, the SMF as a whole is much more stable against different configurations   \citep{Davidzon2013,Ilbert2013}. 
\item The sSFR we derive relies on a differential estimate ($\Delta \log\!\mathcal{M}$) and therefore  systematic errors  (e.g.~due to SED fitting) are expected to cancel out.
\end{enumerate}
%%%%
This last argument holds unless the systematics  vary rapidly as a function of redshift or galaxy type. A comparison with simulated galaxies suggests that this is not the case for our input dataset \citep[Laigle et al.~in preparation, see also][]{Mitchell2013}.  The assumption of a universal IMF is an example of redshift-independent systematics in the SMF  computation. For example, it produces a rigid offset of about $-0.24$\,dex when converting from \citet{Salpeter1955} to \citet{Chabrier2003} IMF \citep[e.g.][]{Santini2015}. 
Another systematic effect, namely the fixed metallicity range use in many  SED fitting codes, is expect to vary slowly with redshift given the evolution of the mass-metallicity relation 
\citep{Sommariva2012,Wuyts2016}. 

We also emphasize that whereas  SMF measurements are usually corrected for the 
\citet{Malmquist1922}, this is rarely quantified in other analyses (e.g., those deriving the sSFR from the MS). The   \citet{Eddington1913} bias is another issue that SMF estimates usually take into account, although the correction technique is still uncertain \citep[see discussion in][]{Davidzon2017}.

%%%%%%%%%%%%%%%% FIG CUMULATIVE MF %%%%%%%%%%%%%%%%%%%
%%%%%%%%%%%%%%%%%%%%%%%% END FIGURE %%%%%%%%%%%%%%%%

\section{Validation tests}
\label{sect:simu}

In this Section we test whether the phenomenological model introduced above is a good description of the sSFR evolution. 
Tests are performed with a semi-analytical model (SAM, Section \ref{sect:simu1}) and hydrodynamical simulations (Section \ref{sect:simu2}). We quantify the impact of the various assumptions on which Equation (\ref{eq:ssfr_ev}) is based, e.g.~the constant SFH. However, before showing the result of our tests, some premises must be clarified.

As the proposed method concerns star forming galaxies, we need to select them in the simulations. After a few experiments, we decided to use their intrinsic sSFR (hereafter sSFR$_0$) as provided by the theoretical model. In fact, 
a cut at $\log(\mathrm{sSFR}_0/\mathrm{yr}^{-1})>-11$ mimics the $NUV-r$ vs $r-J$ classification (NUVrJ) applied to the COSMOS data \citep[][]{Laigle2016}. 
In real surveys, color-color diagrams are preferred to sSFR thresholds since rest-frame colors are more reliable \citep{Conroy2009b} and less SED-dependent \citep{Davidzon2017}.  Theoretically the two classifications are very similar \citep{Arnouts2013} so in the simulation we opted for a simpler sSFR cut. The paucity of quiescent galaxies at $z\gtrsim 3$ makes our results mostly insensitive at this caveat.

Since we want to verify that the framework we built is solid, we do not consider here additional uncertainties such as $z_\mathrm{phot}$ errors and sample incompleteness; moreover, the SMFs we will use in Section \ref{sect:results} have been corrected for this kind of observational biases. A thorough discussion about how to implement observational-like uncertainties in cosmological simulation will be addressed in Laigle et al.~(in preparation).

%%%%%%%%%%%%%%%% FIG SAM %%%%%%%%%%%%%%%%%%%
\begin{figure}
\includegraphics[width=0.99\columnwidth]{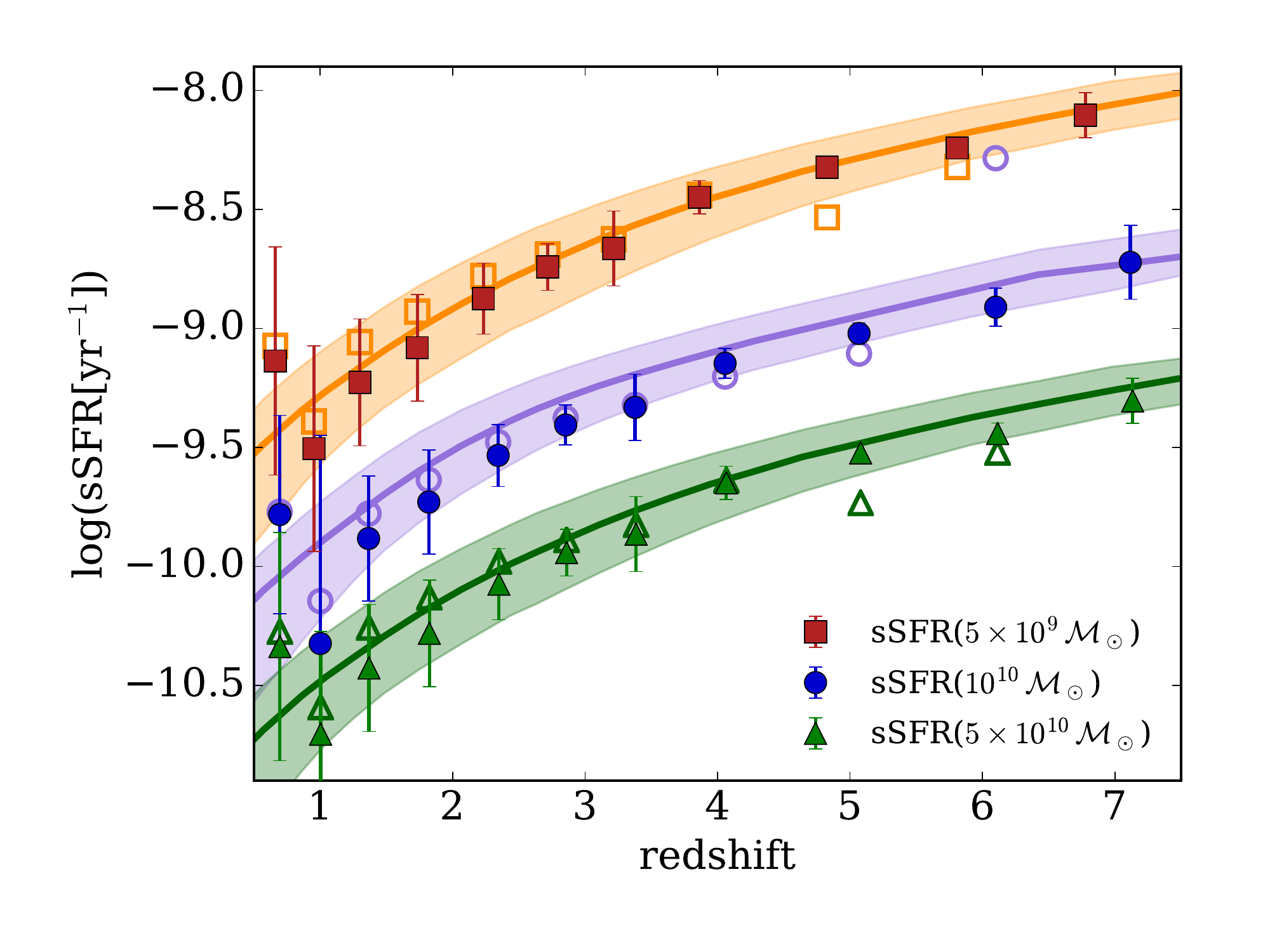}
\caption{Evolution of the sSFR, for galaxies with $\mathcal{M}=5\!-\!50\times10^9\,\mathcal{M}_\odot$ (see legend) in the Munich SAM \citep{Henriques2015}. Filled symbols represent the median of the sSFR($z$) in 20 light-cones (error bars being the standard deviation) derived from their SMF with a constant $\rho_N$ matching. Empty symbols show the results  when we assume an evolving number density \citep[as in][]{Torrey2015}. 
Solid lines are the median sSFR($z$) directly measured from the simulation in the same mass bins, with shaded areas enclosing 16th and 84th percentiles. 
The sSFR at $10^{10}$ and $5\times10^{10}\,\mathcal{M}_\odot$ are shifted downward by 0.6 and 1.2\,dex respectively, to improve readability. } 
\label{fig:ssfr_simu}
\end{figure}
%%%%%%%%%%%%%%%% END FIGURE

\subsection{Test with a semi-analytical model}
\label{sect:simu1}

We verify the reliability of out method by means of cosmological simulations. 
We choose the latest version of the Munich semi-analytical model  \citep[][]{Henriques2015}, based on the Millennium simulation,\footnote{
The Millennium Simulation box has a side of 714\,Mpc when rescaled to the cosmology of \citet{Planck2013-XVI}.} and  
 select 20 independent light-cones with  $1\deg$ diameter aperture to build mock galaxy catalogs similar to  real data. 

For each of these catalogs we estimate the star forming and passive SMFs in several bins of redshift, and trace the $\rho_N$ evolution as described in Section \ref{sect:method3}. 
We then  apply Equation (\ref{eq:ssfr_ev}), setting $f_\mathrm{return}=0.43$ because this is the (instantaneous) mass loss fraction used in the SAM. As we will discuss below, this is a sensitive parameter in our method. 
For $f_\mathrm{merg}$ we use the observational values quoted in Section \ref{sect:method2}  after noticing that they are compatible with the merger rate of Millennium galaxies \citep{Mundy2017}. 

We compare the  sSFR($\mathcal{M}$,$z$) resulting from the SMF evolution with the  median  sSFR$_0$. 
The comparison for three distinct stellar mass bins is shown in Figure~\ref{fig:ssfr_simu}. 
Our method works well between $z\sim2$ and $7$. 
We emphasize that with the abundance matching at constant $\rho_N$ we are able to accurately recover  sSFR$_0$. Following the recipe of \citet{Torrey2015} the results do not change significantly, except at $z\gtrsim5$ where their function produces an additional scatter (Figure~\ref{fig:ssfr_simu}).
This is caused by the small-number statistics of massive halos hosting $\sim10^{10}\,\mathcal{M}_\odot$ galaxies at $z\gtrsim4$   
\citep[we find a similar behavior  using the model of][]{Behroozi2013c}. For the same reason  
\citet{Torrey2015} focus their analysis below $z=3$, where 
the observed universe is more accurately reproduced by 
their hydrodynamical code.   
Supported by these tests, in the following we will link galaxy descendants at constant cumulative number density. 

From the 20 light-cones of \citeauthor{Henriques2015} we can provide a proxy for the cosmic variance expected in observations, since each light-cone has an area similar to the COSMOS field. The statistical error due to cosmic variance is always below $8\%$; this is the scatter in the median sSFR caused by field-to-field variations, not the error affecting the SMF (which propagates into the final outcome, see Section~\ref{sect:discussion1}).   
At $z<1.5$ our estimates are less in agreement with sSFR$_0$, and with larger errors. This mainly depends on the fact that our model has been devised for the early universe. For instance the implemented galaxy merger rate is the one measured at $z\sim2$, and also we did not model star formation quenching at the detailed level required at low $z$ \citep[accounting for environmental effects that modify the SMF shape:][]{Peng2010,Davidzon2016}.

\subsection{Testing systematic effects with an hydrodynamical model}
\label{sect:simu2}

We also test our method using the EAGLE  simulation \citep[][]{Schaye2015}. We take 10 snapshots (from $z=2.01$ to $z=5.97$) of a box with 100 Mpc side.\footnote{
Comoving distance assuming $h=0.6777$, $\Omega_\Lambda=0.693$, $\Omega_\mathrm{m}=0.307$ \citep{Planck2013-XVI}.} 
We adopt the same configuration used in Section \ref{sect:method1}, but with the mass loss fraction  parametrized as in Equation~(\ref{eq:f_return}). In fact, EAGLE code assumes the same IMF of \citet{Henriques2015} but describes $f_\mathrm{return}$ more accurately, as a function of time and metallicity. However, SPS models show that $f_\mathrm{return}$ varies very little from $Z=0.004$ to solar metallicity thus Equation~(\ref{eq:f_return}), which does not include metal enrichment, can be a reasonable approximation of their hydrodynamical model.  

\begin{figure}
\includegraphics[width=0.99\columnwidth]{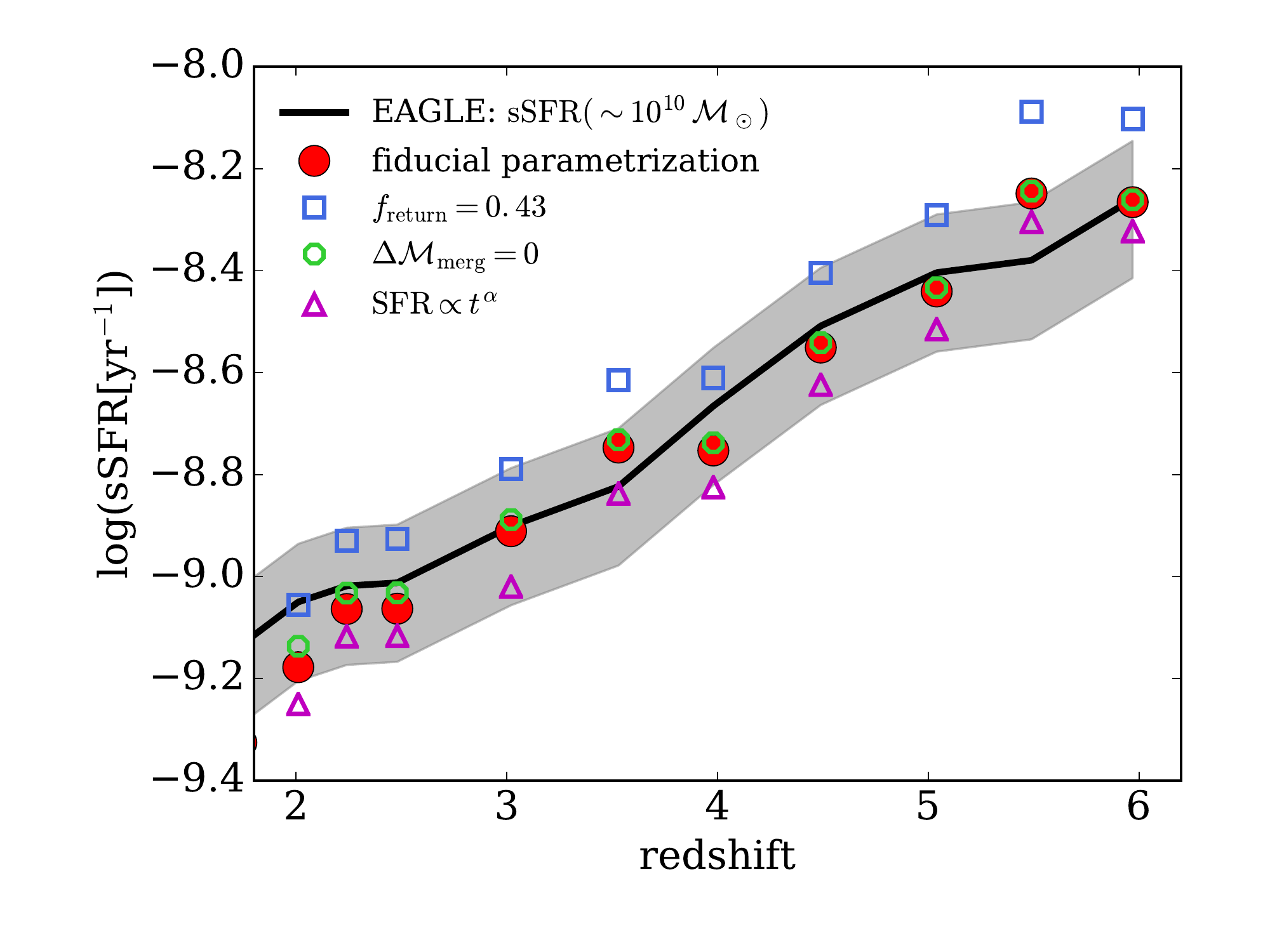}
\includegraphics[width=0.99\columnwidth]{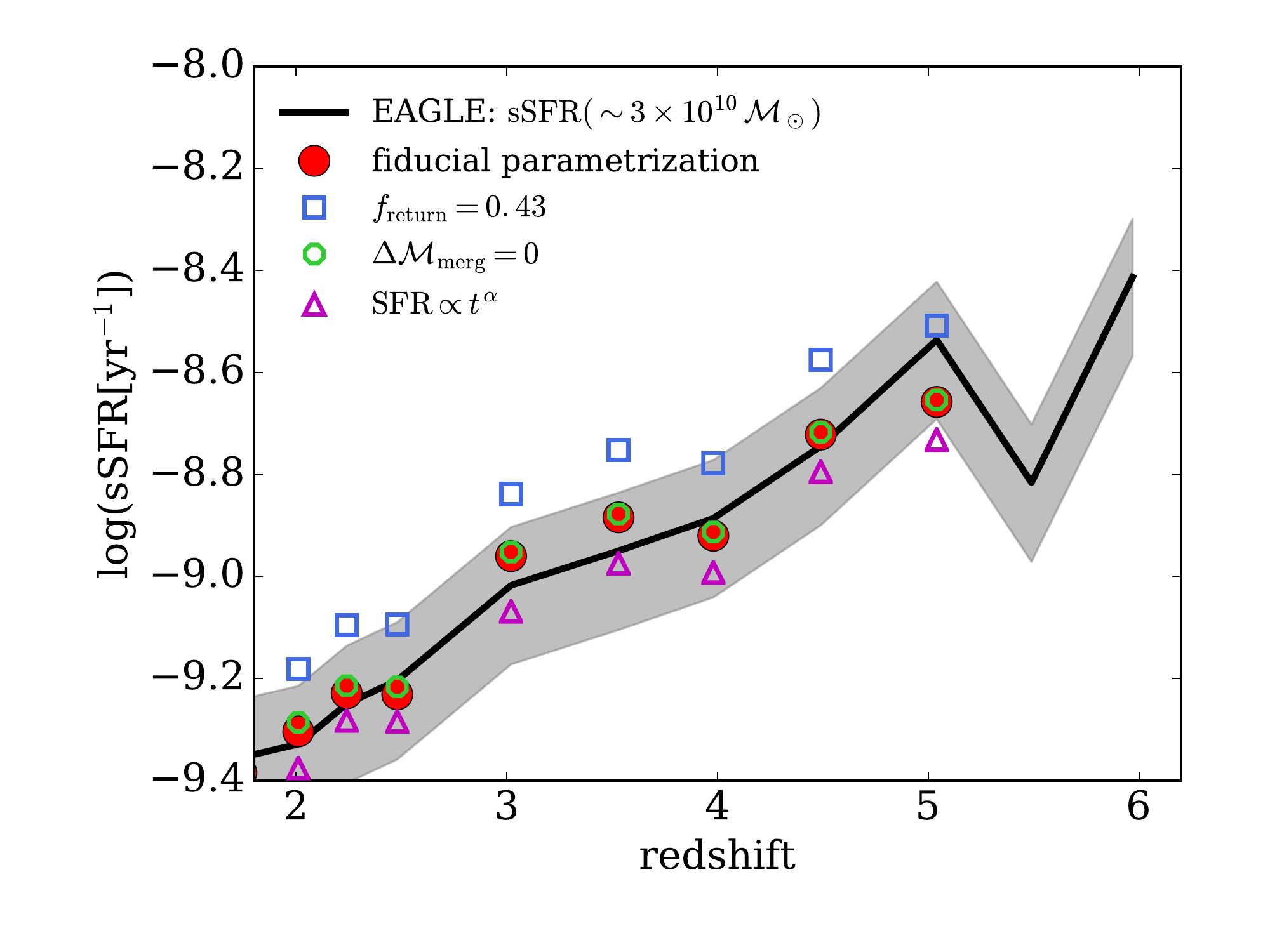}
\caption{Systematic effects related to the main parameters of our method, tested with the EAGLE simulation \citep{Schaye2015}. 
In our ``fiducial configuration'' we assume a  constant SFH with time-dependent $f_\mathrm{return}$,  including the effect of galaxy mergers (see Sect.~\ref{sect:method}). 
The sSFR is computed  at $10^{10}\,\mathcal{M}_\odot$ (\textit{upper panel}) and $3\times10^{10}\,\mathcal{M}_\odot$ (\textit{lower panel}). At larger masses, this test is prevented by the low-number  statistics in the simulated box. Both panels show how the results change if we consider: an instantaneous mass loss fraction $f_\mathrm{return}=0.43$ (blue squares), a negligible mass increase from mergers (green octagons), or a power-law rising SFH (magenta triangles, with $\alpha=1.5$). 
The solid line represents the intrinsic sSFR of EAGLE star-forming galaxies, with a gray shaded area that delimits $\pm30$\% variation.
}
\label{fig:ssfr_hydro}
\end{figure}

In this second test we quantify the systematic effects introduced by our method. Our ``fiducial configuration'' is the one that assumes constant $\mathrm{SFR}(t)$, $f_\mathrm{return}$ as in Equation~(\ref{eq:f_return}), and includes stellar mass assembly via merging (Equation~\ref{eq:m_merg}).  
Given that, we modify each of these parameters separately  (Figure \ref{fig:ssfr_hydro}). To check if there is any $\mathcal{M}$-dependent bias, we perform this test in different mass bins up to $\log(\mathcal{M/M}_\odot)=10.6$, as statistical fluctuation introduce too much noise beyond that threshold. 
Galaxy merging is the one with the smallest impact, as expected from the small value of $\dot{\mathcal{M}}_\mathrm{merg}$ and the short time interval between two $z$-bins. On the other hand, by replacing $f_\mathrm{return}(t)$ with a constant mass-loss fraction \citep[equal to 0.43 for an IMF as in][]{Chabrier2003} the  sSFR increases by $0.10\!-\!0.15$\,dex. 

Systematics related to the SFH, which is kept constant in our fiducial set-up,  are less straightforward to quantify because the choice of different parameterizations is not trivial.  
For instance, exponentially rising SFHs have been proposed as a suitable description of  $z\sim2$  galaxies \citep[][]{Maraston2010} although their rate of star formation ($\propto e^{t_\mathrm{gal}/\tau}$, where $t_\mathrm{gal}$ is the galaxy  age) has been deemed  too extreme \citep{Pacifici2013,Salmon2015}. This is especially true  in our case, since the function in Equation~(\ref{eq:delta_m}) has to reproduce the average SFR. 
In agreement with \citet{Papovich2011} we opt for a power-law function, namely $\mathrm{SFR}\propto t_\mathrm{gal}^{1.5}$ \citep[see also][]{Salmon2015}.  This function needs an additional assumption on the time of  galaxy formation: we parametrize it directly from simulations by fitting the $t_\mathrm{gal}$ distribution of $\sim\!10^{10}\,\mathcal{M}_\odot$ galaxies at different redshifts. 
When the SMF evolution is modeled assuming this power-law SFH, we obtain an sSFR  slightly smaller than the fiducial estimate (Figure \ref{fig:ssfr_hydro}). This trend may be counterintuitive, but we remind that the sSFR is computed at the initial redshift $z_i$. Therefore, with a rising SFH, most of the stellar mass for a given $\Delta\log\mathcal{M}$ will form later.  
However, also with the modified SFR$(t)$, our results lie within $<\!30\%$ from sSFR$_0$ without a strong dependence on stellar mass. A better description of the average SFH should take into account a mix of different  ages and the intrinsic SFR scatter among galaxies with similar mass.  Such a refined characterization is beyond the aims of this paper.

%%%%%%%%%%%%%%%%%%%%% FIG HYDRO
%%%%%%%%%%%%%%%%%% END FIGURE

%%%%%%%%%%%%%%%% SECT RESULTS %%%%%%%%%%%
\section{Observational results}  \label{sect:results}

%%%%%% SHORTCUTS FOR FREQUENTLY CITED PAPERS %%%%%%%
\defcitealias{Davidzon2017}{D+17}
\defcitealias{Grazian2015}{G+15}

We apply our method (with the fiducial configuration) to the observed universe, taking data from \citet[][hereafter D+17]{Davidzon2017}  and \citet[][hereafter G+15]{Grazian2015}. 
The former measured the SMF from the COSMOS2015 galaxy catalog  \citep[][]{Laigle2016} at $z\lesssim6$ over 2\,deg$^2$.\footnote{
This area represents the geometry of the full survey; data used in  \citetalias{Davidzon2017} are restricted to the ``ultra-deep'' stripes after masking saturated stars and corrupted photometric regions (effective area $\sim\!0.62\,\mathrm{deg}^2$).}
\citetalias{Grazian2015} provide an estimate of the SMF between $z=3.5$ and $7.5$, in three combined CANDELS fields (total area $368.90\,\mathrm{arcmin}^2$). 

In both cases we consider their Schechter fits to the $1/V_\mathrm{max}$ points. We use the star-forming SMF of \citetalias{Davidzon2017}  up to $z\sim4$, where the distinction between star forming and passive galaxies is well determined by means of the NUVrJ diagram. In particular, we verify that a redshift-dependent NUVrJ cut evolves too slowly to have an impact on the results \citep[][]{Ilbert2015}. In the star-forming SMF evolution we also have to account for the number density of newly quenched galaxies (see Section \ref{sect:method3}). For the COSMOS2015 sample, this correction factor is derived from the quiescent galaxy fraction ($f_\mathrm{q}$) and its growth with cosmic time. The $f_\mathrm{q}(\mathcal{M},z)$ function is analytically derived  in \citet[][]{Bethermin2017} using COSMOS2015 galaxies as a constraint (see their Equation 2) and is also shown in the inset of Figure \ref{fig:cumu_smf}. 
In CANDELS we work with the total mass function only, assuming that $f_\mathrm{q}$ is negligible at $z>4$.  
We do not connect the CANDELS $z\sim4$ SMF to the COSMOS one at lower $z$, because of the different framework in which they were estimated.

Since we do not have access to the covariance matrix of \citetalias{Grazian2015} Schechter fits, in the $\rho_N$ abundance matching we provide error bars  derived from the Poisson uncertainty of CANDELS galaxy statistics (see Section \ref{sect:discussion1}).  
As mentioned above, we select the Schechter functions from \citetalias{Grazian2015} and  \citetalias{Davidzon2017} because they are corrected for the Eddington bias. We also try the SMF of the ZFOURGE survey \citep[][plot not shown]{Tomczak2014}, although the authors do not take this bias into account. 
Despite some scatter, ZFOURGE presents the same sSFR$(z)$ trend of COSMOS2015, suggesting that most of the Eddington bias may cancel out in our differential estimate \citep[or confirming that the medium-band imaging of ZFOURGE results in smaller $z_\mathrm{phot}$ and $\mathcal{M}$ errors]{Straatman2016}.

\begin{figure*}
\centering
\includegraphics[width=0.75\textwidth]{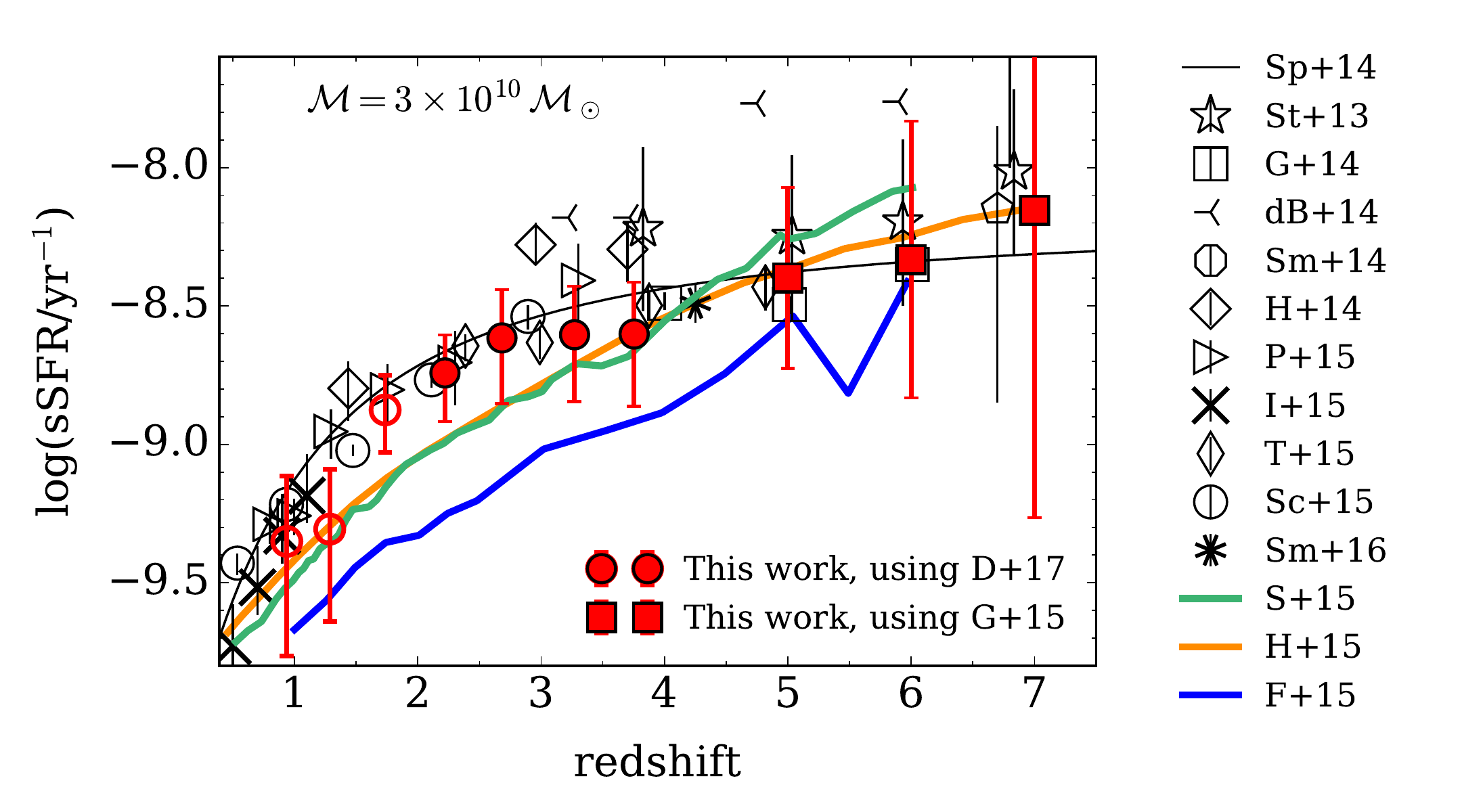}
\caption{The sSFR at $3\times10^{10}\,\mathcal{M}_\odot$ 
derived from the SMF evolution as observed in \citet[][red circles]{Davidzon2017} and 
\citet[][red squares]{Grazian2015}. Although the method is optimized for $z>2$, we also show our results at lower redshifts (empty red circles). 
Estimates from Fig.~\ref{fig:ssfr_lit} are reported here using the same symbols. In addition, we show the sSFR computed in 
\citet[][]{Ilbert2015}, \citet[][]{Schreiber2015}, \citet[][]{Pannella2015}, \citet[][]{Tasca2015}, \citet[][]{Heinis2014}, \citet[][]{Smit2016}.  
The black solid line is the \citet{Speagle2014} fitting function  calculated at  $3\times10^{10}\,\mathcal{M}_\odot$.
Blue, green, and orange lines show the sSFR($z$) from  EAGLE, Illustris, and the latest Munich simulation respectively  \citep{Furlong2015,Sparre2015,Henriques2015}.}
\label{fig:ssfr_result}
\end{figure*}

We compute the sSFR($\mathcal{M},z$) at several fixed values of  stellar mass (Table \ref{tab:ssfr_result}). 
Our results at $\log(\mathcal{M/M}_\odot)\simeq10.5$, close to the characteristic mass $\mathcal{M}_\star$, are shown in Figure~\ref{fig:ssfr_result}. For sake of completeness we  also include our estimates at $z<2$, despite the fact that the parameters of our method are calibrated for higher redshifts (Section~\ref{sect:method2}); errors at such a low redshift are larger than $z=2\!-\!4$ mainly because $f_\mathrm{q}$ is higher, and its uncertainty significantly contributes to the total error budget. 
At $z>2$ we find a shallow  sSFR evolution, proportional to $(1+z)^{1.1\pm 0.2}$. This trend is similar  to what found in \citet{Gonzalez2014} and \citet{Tasca2015}, while other sSFR estimates  are higher in normalization \citep[e.g.][]{Heinis2014} or steeper in their redshift evolution \citep[e.g.,~$\propto\!(1+z)^{1.5}$ in][]{Faisst2016a}. 
However, the  stellar mass range in one study may significantly differ from the others, making the comparison less straightforward especially if the  linearity between 
$\log(\mathrm{SFR})$ and $\log(\mathcal{M})$ breaks up
 \citep{Whitaker2014}. 
To avoid confusion, in Figure~\ref{fig:ssfr_result} we show only analyses with   median stellar masses  comparable to ours (i.e., it does not exceed a factor $\sim\!3$ difference). 
We make an exception for studies at $z>4$, since none of them  effectively probe  $\mathcal{M}>10^{10}\,\mathcal{M}_\odot$  \citep{Stark2013,Gonzalez2014,deBarros2014b,Smit2016}. This is indeed a distinctive feature of our method, which is effectively also in the highest-mass regime if the exponential tail of the SMF is sufficiently well constrained.

%%%%%%%%%%%%%%%% FIG RESULTS %%%%%%%%%%%%%%%%%%%
%%%%%%%%%%%%%%%%%%%%%%%%%%%%%%%%%%%%%%%%%

\section{Discussion}
\label{sect:discussion}

\subsection{Selection effects and error budget}
\label{sect:discussion1}

Despite consistent within $1\sigma$, 
our estimates  are slightly   below  the 
``concordance'' sSFR function    
resulting from the comprehensive review of \citet[][]{Speagle2014}.\footnote{
\citeauthor{Speagle2014} combine coherently MS data from 25 different studies, to which they fit the functional form $\log\mathrm{SFR}(t,\log\mathcal{M})=(a_1 t + a_2)\log\mathcal{M} + (b_1 t +b_2)$. }
We argue that the offset can be ascribed to a different galaxy selection between our analysis and those  in \citeauthor[][]{Speagle2014}: 
Using the NUVrJ diagram, our star-forming class includes also galaxies with moderate SFRs, which instead fall in the passive locus when using $U-V$ vs $V-J$ \citep[UVJ, see discussion in][]{Muzzin2013b}. 
On the other hand, several studies comprised in \citeauthor{Speagle2014} may be biased towards bluer galaxies,  due to their selection technique (e.g., LBG criteria) or because they intentionally focus on the core of the MS \citep[e.g., by applying a $\sigma$-clipping to the distribution,][]{Santini2017}.
Interestingly, the $z<1.4$ estimates from \citet{Ilbert2015}, in which the SFRs are derived from UV-IR balance,  also lie systematically below the concordance sSFR function. 
Based on a classification similar to ours, they are consistent with our trend   (Figure \ref{fig:ssfr_result}). 

Another potential problem is related to heavily dust-attenuated starburst galaxies, which may be missing in our sample. Thanks to the COSMOS2015 panchromatic detection strategy, which 
results in a high completeness of our sample, 
we conclude that such a bias is negligible   \citepalias[\citealp{Laigle2016};][]{Davidzon2017}. 
This is confirmed by the good agreement with the sSFR of \citet[][]{Schreiber2015}, derived with a different technique (far-IR stacking of \textit{Herschel} images) but also based on a mass-complete galaxy catalog. 
On the other hand, recent observations indicate that the dust content  in high-$z$ galaxies varies over a wide range \citep[e.g.,][]{Faisst2017b}, which  is a major concern for the SFR estimates based on rest-frame UV luminosity and has to be investigated in more detail with future far-IR measurements.

%%%%%%%%%%%%%%%%%%%%%%% FIGURE  
\begin{figure}
\includegraphics[width=0.99\columnwidth]{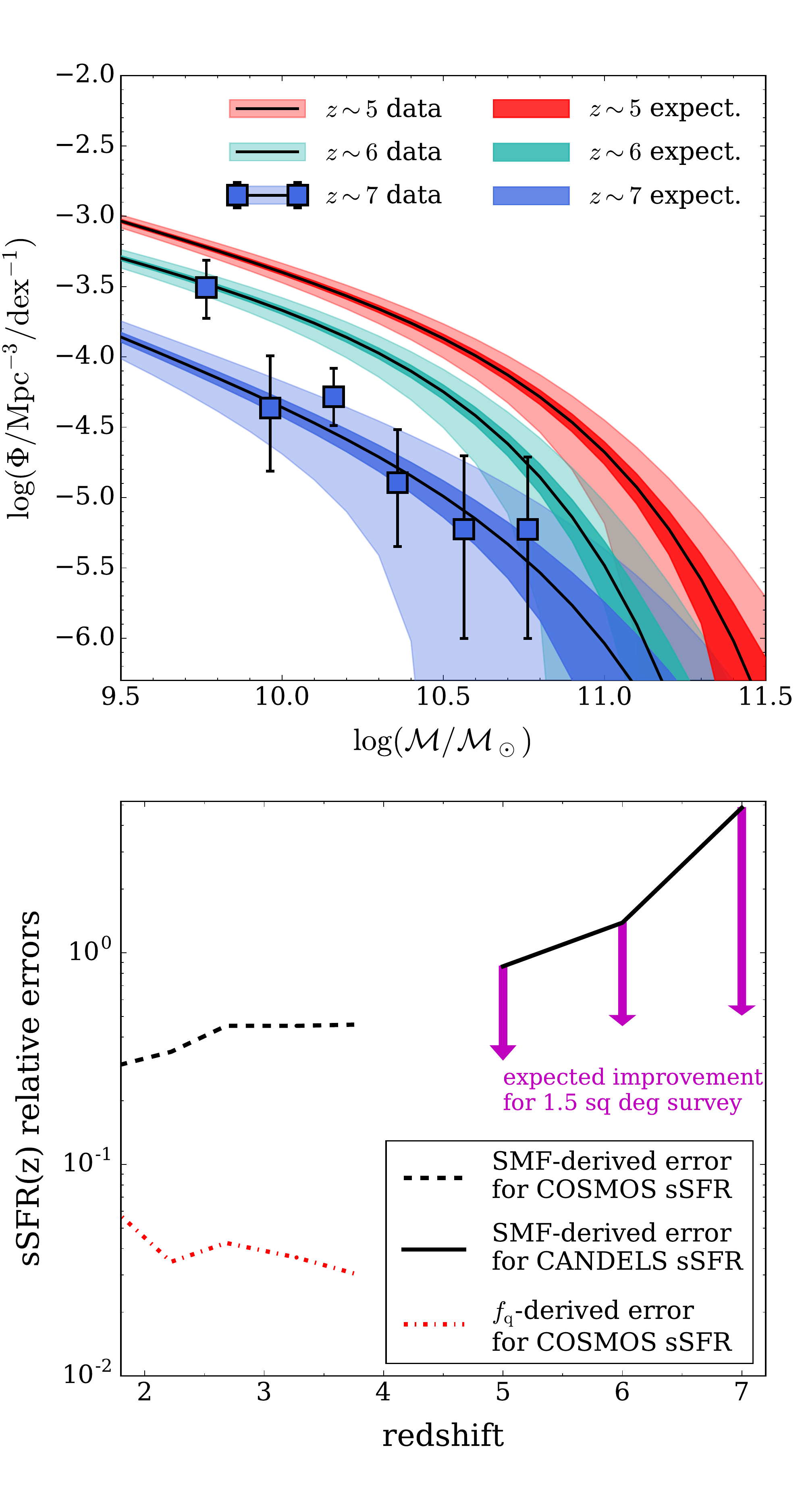}
\caption{\textit{Upper panel:} The SMF of CANDELS galaxies between $z=5$ and 7  \citetalias[black lines, from][]{Grazian2015}. For each Schechter function, a transparent area shaded in light color encloses the $1\sigma$ uncertainty of the fit, calculated by taking into account Poisson noise only. At $z=7$, the $1/V_\mathrm{max}$ estimates are also plotted (blue squares). 
Opaque shaded areas (in darker colors) show the reduced statistical error that are expected for a CANDELS-like survey over 1.5\,deg$^2$.
\textit{Lower panel:} Relative errors $\sigma_\mathrm{sSFR}/\mathrm{sSFR}$, with $\sigma_\mathrm{sSFR}$ corresponding to half the error bar of our estimates shown in Figure \ref{fig:ssfr_result}.  At $z<4$ we decompose two sources of uncertainty: the error of the star-forming SMF (dashed line) and the one of the quiescent fraction used to correct for recently quenched galaxies (dot-dashed line, see Section~\ref{sect:discussion1}). 
At $z>4$ only the former is considered (solid line). 
Arrows indicate the expected error reduction at $z>4$, if the SMF were constrained by a 1.5\,deg$^2$ survey instead of the CANDELS data used in \citetalias{Grazian2015}.    
}
\label{fig:errors}
\end{figure}
%%%%%%%%%%%%%%%%%%%%%%%%%%%%%

For the sSFR derived from the \citetalias{Davidzon2017} SMF evolution, error bars include the $1\sigma$  uncertainty of the Schechter function along with the one of $f_\mathrm{q}(\mathcal{M},z)$. The former is the dominant source of uncertainty (Figure~\ref{fig:errors}, bottom panel). The Schechter function in \citetalias{Davidzon2017} is a fit to the $1/V_\mathrm{max}$ determinations taking into account  
Poisson noise, cosmic variance, and the scatter due to SED fitting uncertainties. 
The resulting sSFR precision is comparable with the one of other  studies, although some of them are reported in Figure~\ref{fig:ssfr_result}  with the smaller errors derived e.g.~from a stacking procedure, rather than the variance of the measurements. For instance, for \citet{Tasca2015} we plot the error on the median ($\sigma_\mathrm{SFR}/\sqrt{N}$) but the authors find  $\sigma_\mathrm{SFR}=0.2\!-\!0.4$\,dex when they consider the $\log(\mathrm{sSFR})$ dispersion. Error bars in \citet{Schreiber2015} are of the order of 0.1\,dex  but they do not include the effect of $z_\mathrm{phot}$ and $\mathcal{M}$ uncertainties (the authors also state $\sigma_\mathrm{SFR}=0.3$\,dex).   

At $z>4$ the SMF measurements become more uncertain, especially at $z\sim7$, which is reflected in the sSFR. 
As noted in Section~\ref{sect:results}, \citetalias{Grazian2015} provide only the marginalized $1\sigma$ error for each Schechter parameter. Without a covariance matrix, we approximate the $1\sigma$ error of the best-fit Schechter function deriving it from the Poisson statistics in the CANDELS volume (Figure~\ref{fig:errors}, upper panel).
Although in this way the SMF uncertainty is likely underestimated, the relative error $\sigma_\mathrm{sSFR}/\mathrm{sSFR}$ (Figure~\ref{fig:errors}, bottom panel) is nonetheless large, up to a factor $\sim\!3$ at $z=7$. This illustrates the impact of small-number statistics in current high-$z$ surveys. 

New observations over 1\,deg$^2$, with resolution and depth similar to the CANDELS wide fields, should drastically reduce this issue. Combining them  with existent CANDELS data, one would cover about $1.5$\,deg$^2$, with the 1\,deg$^2$ contiguous area probing cosmic structures over larger scales.   In Figure~\ref{fig:errors} (upper panel) we show how the Poisson noise would decrease in the SMF of this hypothetical  $1.5$\,deg$^2$ galaxy survey, omitting the additional improvement in terms of cosmic variance. The expected gain in $\sigma_\mathrm{sSFR}/\mathrm{sSFR}$ is shown by arrows in the lower panel of the figure. 
We argue that the \textit{Hubble} Space Telescope (HST) is the most suitable facility to achieve this goal. The 
\textit{James Webb} Space Telescope (JWST) is not ideal for such a large-area survey, because overheads may reach $\sim\!80\%$ the exposure time (according to the JWST planning tool).  
Moreover,  HST can provide parallel observations in optical bands, a unique benefit for supplementary $z<3$ studies. JWST instruments may then be used with a follow-up strategy, e.g.~to better calibrate SED fitting estimates.

\subsection{The massive end of the star-forming MS}
\label{sect:discussion2}

\begin{figure}
\includegraphics[width=0.99\columnwidth]{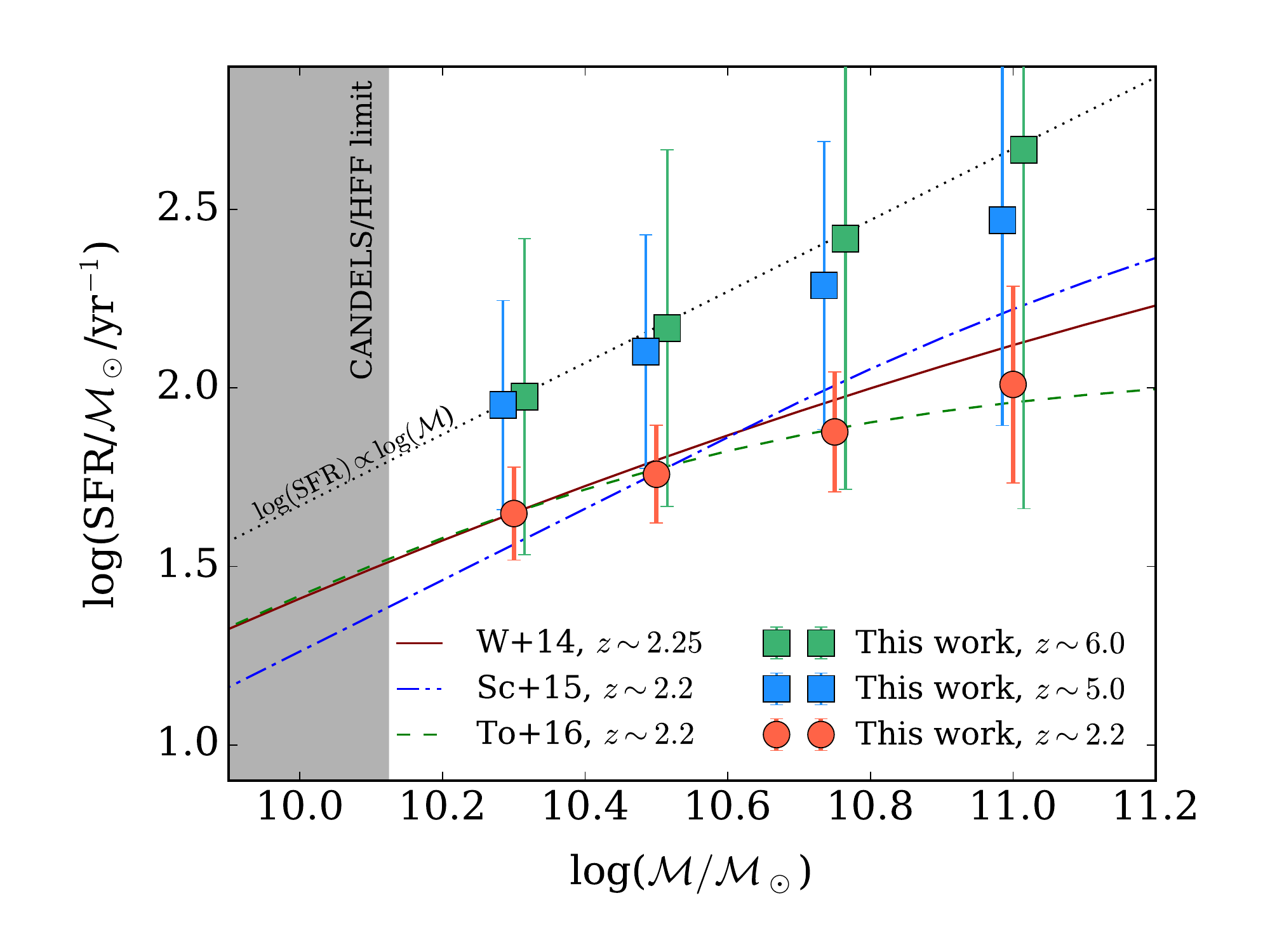}
\caption{First constraints to the star-forming MS by using the 
sSFR$(\mathcal{M},z)$ derived from the SMF evolution. 
Estimates relying on \citetalias{Davidzon2017} are shown with filled circles, while squares indicated when the \citetalias{Grazian2015} SMF is used (an horizontal offset of $\pm0.015$\,dex is applied to improve readability). 
Colored lines are not fits to the points, but the SFR$(\mathcal{M})$ functions calibrated in \citet[][solid line]{Whitaker2014}, \citet[][dot-dashed line]{Schreiber2015}, and \citet[][dashed line]{Tomczak2016} at the same redshift.  A shaded area delimits the maximal stellar mass  in other studies probing the MS at $z>3$ \citep{Salmon2016,Santini2017}.
 }
 \label{fig:ms_turnoff}
\end{figure}

The dependency of the SFR on stellar mass is also illustrated in Figure~\ref{fig:ms_turnoff}, where we multiplied  sSFR$(\mathcal{M},z)$  by $\mathcal{M}$ to plot the star-forming MS at $\log(\mathcal{M/M}_\odot)>10.2$. This is the mass threshold beyond which \citet{Whitaker2014} find a  flattening of the $\mathrm{SFR}$-$\mathcal{M}$ relation at $z<2.5$ \citep[see also][]{Ilbert2015,Lee2015,Schreiber2015}. 
 Our $z<3$ data also suggest this trend. Figure~\ref{fig:ms_turnoff} shows the MS we obtain at $z\sim2.2$,  along with the functional forms determined in \citet{Whitaker2014},  \citet{Schreiber2015}, and \citet{Tomczak2016}. The difference among the three studies is mainly due to the way their samples are built. 

A turnover mass is observed by \citep{Tomczak2016} up to $z\sim4$, although recently the ALMA Redshift 4 Survey has found opposite results \citep{Schreiber2017}.
Besides them, other MS studies at high redshift probe a less massive regime:   CANDELS \citep[see][]{Salmon2015} or  the HST Frontier Field \citep[HFF,][]{Santini2017}  do not have enough statistical power at $\log(\mathcal{M/M}_\odot)>10.2$. 
For similar reasons, the high-mass end of the $z>4$ SMF is not well constrained and the MS we derive is highly uncertain.   
However, the SMF accuracy is easier to enhance than the statistics in SFR measurements, which  require additional far-IR or sub-mm data. In this perspective, our method is expected to be an effective tool to constrain the massive end of the MS. 
At present, our analysis barely suggests that there is a flattening in the MS fit  already at $z\sim5$, while at higher redshift the relation $\mathrm{SFR}\propto\mathcal{M}^\alpha$  holds also for the most massive galaxies (with $\alpha$ close to unity as in the local universe, see Figure~\ref{fig:ms_turnoff}). 
If confirmed, this preliminary result will put fundamental constraints on the quenching time-scales of massive galaxies in the early universe, and their role during the epoch of reionization \citep[cf.][]{Sharma2016}.

%%%%%%%%%%%%%%%%% FIGURE MAIN SEQ
%%%%%%%%%%%%%%%% END FIGURE

\subsection{Comparison to simulations}
\label{sect:discussion3}

In Figure~\ref{fig:ssfr_result} we compare our results to state-of-the-art semi-analytical and hydrodynamical models   \citep{Sparre2015,Henriques2015,Furlong2015}. These simulations suggest that the sSFR at $z<3$ is  $\propto(1+z)^{2\!-\!2.5}$ as 
, while we find a shallower increase (with exponent $\sim1.1$).

\begin{figure}
\includegraphics[width=0.99\columnwidth]{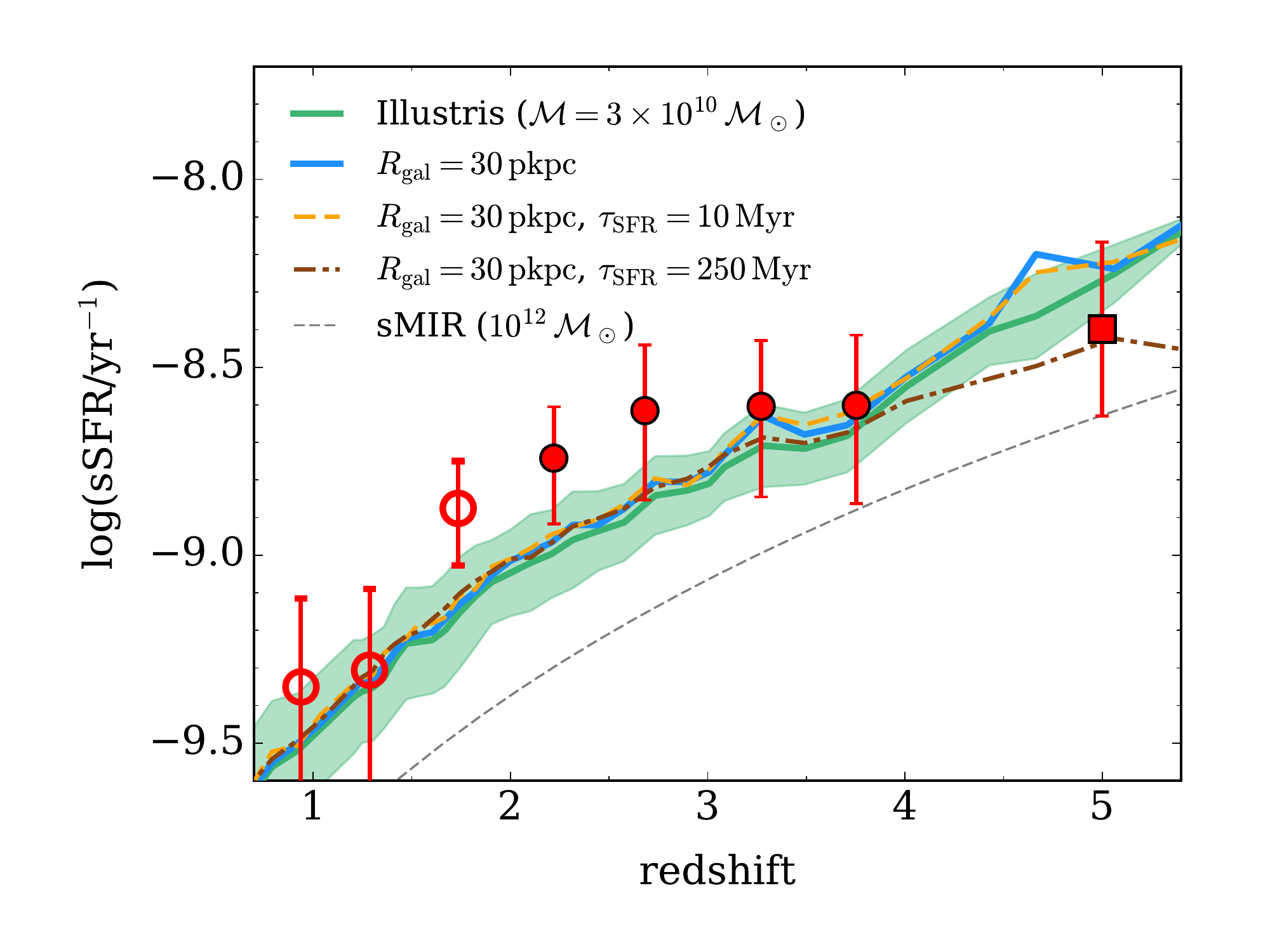}
\caption{Comparison between our results (red symbols with error bars, as shown in Figure~\ref{fig:ssfr_result}) and the sSFR predicted by Illustris at $3\times10^{10}\,\mathcal{M}_\odot$. The solid green line and its shaded area are median sSFR  and rms from Illustris,  computed similarly to  \citet[][]{Sparre2015} but  selecting only galaxies with $\mathrm{sSFR}>10^{-11}\,\mathrm{yr}^{-1}$.  
Furthermore, stellar mass and SFR of simulated galaxies have been 	progressively modified by  considering: a galaxy radius ($R_\mathrm{gal}$) matching the one probed by aperture-corrected photometry (solid  blue line) and the SFR over the last 10 or 250\,Myr (dashed yellow and dot-dashed brown lines). The short-dashed gray line is the specific (dark) matter increase rate \citep[sMIR,][]{Neistein&Dekel2008} calculated for the average mass of the hosting halos as a function of redshift ($\sim\!10^{12}\,\mathcal{M}_\odot$).
}
\label{fig:ssfr_results_simu}
\end{figure}

We first note that the slope is the same for the three simulations, although EAGLE has a lower normalization suggesting that the implemented stellar feedback could be too strong in its case  \citep[][]{Furlong2015}. 
This similarity is likely due to the fact that the star formation is tightly connected to the underlying dark matter accretion, irrespective of the different sub-grid baryon physics coded in the three simulations.   
To illustrate this point, in Figure~\ref{fig:ssfr_results_simu} we contrast the sSFR of Illustris galaxies at $\log(\mathcal{M/M}_\odot)\simeq10.5$  to the specific dark matter accretion rate of their $\sim\!10^{12}\,\mathcal{M}_\odot$ hosting halos \citep[see also][]{Weinmann2011,Lilly2013}.

Another reason for the discrepancy may be the presence of some selection effect.
Previous comparisons between models and data have often 
classified galaxies through inconsistent criteria.  
For instance, \citet{Dave2016} extract from their $z>1$ simulation all the galaxies with $\log(\mathrm{sSFR/yr}^{-1})>-10.9$ and compare them to \citet{Whitaker2014}, whose analysis based on the UVJ selection tends to exclude  $\mathrm{sSFR\sim10^{-10}\,\mathrm{yr}^{-1}}$ galaxies from the star-forming class \citepalias[see][]{Davidzon2017}. We also mention \citet{Sparre2015}, where the sSFR of all the Illustris galaxies in a given mass bin is compared to the sSFR function of \citet{Behroozi2013b}, i.e.~a fit to star-forming galaxies only. 
This choice however has no influence on their findings owing to the scarcity of massive quiescent galaxies in Illustris.

In our work the simulated star-forming sample is designed to be consistent to the observed one. 
The theoretical predictions shown in Figure~\ref{fig:ssfr_result} come from sSFR$_0$, i.e.~the median of simulated galaxies after applying a cut at $\mathrm{sSFR}>10^{-11}\,\mathrm{yr}^{-1}$. This is a good proxy of the NUVrJ classification used in  \citetalias{Davidzon2017}.\footnote{  
See also Sect.~\ref{sect:simu1}. 
The alternative of using the NUVrJ  diagram in both cases  is not convenient, as there may be substantial differences in the computation of rest-frame magnitudes  \citep[see][for tests on the UVJ diagram]{Henriques2015}.}
We also verify that neither a conservative cut at $10^{-9.5}\,\mathrm{yr}^{-1}$ can reconcile the lower sSFR in the simulations, implying that it is not caused by an excess of post-starburst galaxies \citep[i.e., an over-populated ``green valley'' due to a too long quenching time-scale,][]{Moutard2016b}. 

Other observational biases  that can impair the comparison are related to the way stellar mass and SFR are estimated. Hydrodynamical models usually define $\mathcal{M}$ as the sum of stellar particles gravitationally bounded to the galaxy sub-halo. This is an overestimate of the SED-fitting stellar mass, which is derived from aperture-corrected photometry and  does not take into account the galaxy outskirt (i.e., the intra-cluster light). 
A value of sSFR closer to the observed one is obtained by considering only the inner 30 physical kpc of the given sub-halo \citep{Schaye2015}. 

Regarding galaxy star formation, the instantaneous SFR should be replaced with an estimate averaged over the last $10\!-\!200$\,Myr, i.e.~the time-scales probed by the main SFR indicators  \citep{Sparre2017a}.  
Moreover, the SPS model used to build the SED templates may assume a stellar mass loss significantly different from the one of the simulation.\footnote{
This caveat must be kept in mind also when comparing studies that have the same IMF: although the initial abundance of low-mass stars is the same,  winds during the Asymptotic Giant Branch phase  may be modeled in a different way (Laigle et al.~in preparation). }

%%%%%%%%%%%%% FIGURE

%%%%%%%%%%% END FIGURE

To study the impact of these biases on our comparison, we modify the following properties of Illustris galaxies:
%%%%%
\begin{enumerate}[1)]
\item we re-compute $\mathcal{M}$ and SFR by including only stellar particles within 30 physical kpc from the galaxy center (the density peak of the sub-halo); \label{item:mstar_30}
\item tracing backward the SFH of each particle, we calculate the SFR on two different time-scales:  $\tau_\mathrm{SFR}=10$ and 250\,Myr; \label{item:sfr_myr}
\item we modify each stellar particle by replacing their original return fraction with the one that follows Equation (\ref{eq:f_return}). \label{item:f_ret}
\end{enumerate}
%%%%%

The outcomes of such a post-processing are summarized in Figure~\ref{fig:ssfr_results_simu}. The $f_\mathrm{return}$ modification (step \ref{item:f_ret}) has a negligible impact and therefore is not included in the figure. This confirms our findings in Section \ref{sect:simu2}, where we showed that Equation (\ref{eq:f_return}) is a good approximation of the return fraction implemented in the EAGLE code. 
The exclusion of the sub-halo outskirts (step \ref{item:mstar_30}) and the longer SFR time-scales (\ref{item:sfr_myr}) are not able to reconcile the $z<3$ theoretical predictions with our results. At higher redshifts a better agreement is reached when the sSFR of Illustris galaxies is calculated over the last 250\,Myr, but the small-number statistics at $z>5$ prevent us from drawing strong conclusions. 

If observational biases cannot account for the $z<3$ discrepancy, a solution should come from some improvement in the model's physical prescriptions,  likely concerning black-hole (BH) driven winds given the mass range we probe.  
In the ``next generation''  of Illustris (Illustris-TNG) 
 both galactic winds and BH feedback are indeed modified  \citep[along with numerous other parameters, see][]{Weinberger2017,Pillepich2017} and significant changes are expected especially for the sSFR of galaxies at $\log(\mathcal{M/M}_\odot)>10.2$.

%%%%%%%%%%%%%%%%%%% TABLE
\begin{table*}
\caption{Specific SFRs derived from the SMF evolution, for star-forming galaxies between $2\times10^{10}$ 
and $10^{11}\,\mathcal{M}_\odot$ in different redshift bins. Null value is given in the range of stellar mass and redshift dominated by quiescent galaxies. }
\label{tab:ssfr_result}
\setlength{\extrarowheight}{1ex}
\centering
\begin{tabular}{lccccc}
\hline\hline
redshift & $\langle z \rangle^\dagger$ & \multicolumn{4}{c}{$\mathrm{\log(\mathrm{sSFR/yr}^{-1})}$}  \\ 
 & & at $\log(\mathcal{M/M}_\odot)=10.3$ & $\log(\mathcal{M/M}_\odot)=10.5$ & $\log(\mathcal{M/M}_\odot)=10.7$ &  $\log(\mathcal{M/M}_\odot)=11.0$  \\
[1ex] \hline 
%%%% start  
& & \multicolumn{4}{c}{using \citet{Davidzon2017} SMF:}  \\
 $0.8<z\leqslant1.1$ & $ 0.94^\ast $ & $-9.40^{+0.26 }_{-0.49 }$ & $-9.33^{+0.23 }_{-0.38 }$   & $-9.25^{+0.19 }_{-0.28 }$ & $\dots$ \\
 $1.1<z\leqslant1.5$ & $ 1.29^\ast $ & $-9.20^{+0.20 }_{-0.28 }$ & $-9.29^{+0.21 }_{-0.31 }$   & $-9.51^{+0.27 }_{-0.61 }$ & $\dots$ \\
 $1.5<z\leqslant2.0$ & $ 1.74^\ast $ & $-8.87^{+0.14 }_{-0.16 }$ & $-8.87^{+0.12 }_{-0.15 }$   & $-8.87^{+0.13 }_{-0.15 }$ & $\dots$ \\
 $2.0<z\leqslant2.5$ & $ 2.22 $ & $-8.65^{+0.13 }_{-0.15 }$ & $-8.74^{+0.14 }_{-0.17 }$   & $-8.85^{+0.16 }_{-0.20 }$ & $-8.99^{+0.27 }_{-0.38 }$ \\
 $2.5<z\leqslant3.0$ & $ 2.68 $ & $-8.59^{+0.18 }_{-0.24 }$ & $-8.61^{+0.17 }_{-0.24 }$   & $-8.70^{+0.20 }_{-0.26 }$ & $-8.82^{+0.30 }_{-0.41 }$ \\
 $3.0<z\leqslant3.5$ & $ 3.27 $ & $-8.62^{+0.19 }_{-0.24 }$ & $-8.60^{+0.17 }_{-0.24 }$   & $-8.65^{+0.20 }_{-0.25 }$ & $-8.84^{+0.35 }_{-0.47 }$ \\
 $3.5<z\leqslant4.0$ & $ 3.75 $ & $-8.56^{+0.16 }_{-0.21 }$ & $-8.60^{+0.19 }_{-0.26 }$   & $-8.64^{+0.24 }_{-0.35 }$ & $-8.71^{+0.36 }_{-0.62 }$ \\
& & \multicolumn{4}{c}{using \citet{Grazian2015} SMF:}  \\
 $4.5<z\leqslant5.5$ & $ 5 $  & $-8.35^{+0.29 }_{-0.37 }$  & $-8.40^{+0.32 }_{-0.43 }$ & $-8.45^{+0.36 }_{-0.53 }$ & $-8.53^{+0.47 }_{-0.86 }$ \\
 $5.5<z\leqslant6.5$ & $ 6 $  & $-8.32^{+0.43 }_{-0.88 }$  & $-8.33^{+0.46 }_{-1.07 }$ & $-8.34^{+0.51 }_{-1.46 }$ & $-8.33^{+0.77 }_{-2.58 }$ \\
 $6.5<z\leqslant7.5$ & $ 7 $  & $-8.01^{+0.57 }_{-0.81 }$  & $-8.15^{+0.91 }_{-2.00 }$ & $-8.33^{+0.90 }_{-2.44 }$ & $-8.69^{+0.97 }_{-2.07 }$ \\
%%% end  
[1ex] \hline 
\end{tabular}
\begin{flushleft}
\begin{footnotesize}
$^\dagger$For \citetalias{Davidzon2017}, $\langle z\rangle$ is the median $z_\mathrm{phot}$ of galaxies in the given $z$-bin; for \citetalias{Grazian2015} it is  the center of the bin.  
$^\ast$Estimates below the optimal redshift range of the method. 
\end{footnotesize}
\end{flushleft}
\end{table*}

\section{Conclusions}
\label{sect:conclusions}

Thanks to deeper observations in the COSMOS and CANDELS fields, recent work  has measured the $2<z<7$ SMF with unprecedented accuracy \citep{Davidzon2017,Grazian2015}. 
On this premise, we devised an advanced version of the semi-empirical technique described in  \citet{Ilbert2013}, to track the SMF evolution of star-forming galaxies and derive their median sSFR as a function of redshift.  
In contrast to other studies that require a direct measurement of the SFR, the fundamental ingredients here are  $z_\mathrm{phot}$ and $\mathcal{M}$ estimates, along with a reliable  technique (the NUVrJ diagram) to distinguish between star forming and passive galaxies. 
For this reason our method does not require expensive far-IR or sub-mm observations to derive SFRs, but only robust photometry in optical and near-IR bands.  
Moreover, issues like Malmquist and Eddington biases are routinely tackled in SMF analyses, while for other sSFR measurements it is more difficult to quantify their impact. 
However, we are limited to probing stellar masses  $>\!10^{10}\,\mathcal{M}_\odot$ because the highest $z$-bins start to be incomplete below this threshold. 
\vspace{10pt}

Using mock galaxy catalogs from cosmological simulations we demonstrated that our method works remarkably well from $z\sim2$ up to $z\sim7$. Most of the assumptions (e.g.~related to SFH or galaxy merging) do not introduce significant systematics ($<\!30\%$). The parametrization of stellar mass loss must be carefully modeled, because a simplistic instantaneous return fraction can change the sSFR by $\sim\!0.1$\,dex or more. The return fraction we assumed (Eq.~\ref{eq:f_return}) is in agreement with the one implemented in hydrodynamical simulations (EAGLE, Illustris). 
\vspace{10pt}

Using the  SMFs observed in COSMOS and CANDELS  we found that $\mathrm{sSFR}\propto (1+z)^{1.1\pm0.2}$ at $2<z<7$.  
At $z<4$ the trend of our data is in marginal tension with the theoretical expectation, even after correcting for the different $\mathcal{M}$ and SFR definitions (e.g., the time-scale over which the simulated SFR is calculated). This suggests a revision of the sub-grid physics in the models. 
\vspace{10pt}

This work is preparatory to exploiting next-generation SMF estimates. 
At present, some of our findings are affected by the large uncertainties at $z>4$ (which are however of the same order of magnitude of other studies). One of these tentative results is the determination of the MS high-mass end from the observed sSFR$(z)$ up to $z\sim6$, to identify the epoch when the linear relation between $\log(\mathrm{SFR})$ and $\log(\mathcal{M})$ starts to flatten. 
Future missions like \textit{Euclid} or the Wide-Field Infrared Survey Telescope (WFIRST) will give the opportunity to significantly improve the galaxy SMF, especially by combining their data with the IR photometry from the Euclid/WFIRST Spitzer Legacy Survey (PI: P.~Capak). 
Moreover, we have shown that if future HST observations covered a total area of $\geqslant\!1.5\,\mathrm{deg}^2$ with a depth similar to the CANDELS wide-fields, the statistical errors would dramatically reduce. Then, our technique shall provide unique constraints to the sSFR and MS evolution of massive galaxies.

\acknowledgements
We thank  Paola Santini for providing their data in a convenient digital format, and Bahram Mobasher for his useful comments.
%Spitzer
This work is based on observations and archival data made with the Spitzer Space Telescope, which is operated by the Jet Propulsion Laboratory, California Institute of Technology under a contract with NASA. Support for this work was provided by NASA.
%IRSA
This research has made use of the NASA/IPAC Infrared Science Archive, which is operated by the Jet Propulsion Laboratory, California Institute of Technology, under contract with the National Aeronautics and Space Administration.
% HSC:
Also based on data collected at the Subaru Telescope and retrieved from the HSC data archive system, which is operated by Subaru Telescope and Astronomy Data Center at National Astronomical Observatory of Japan
OI acknowledges funding of the French Agence Nationale de la Recherche for the SAGACE project. We remark the financial support the COSMOS team receives from the Centre National d'\'{E}tudes Spatiales.

\end{document}